# Highly-efficient third-harmonic generation from ultrapure diamond crystals

AIZITIAILI ABULIKEMU[⊥] AND MUNEAKI HASE[*]

*Department of Applied Physics, University of Tsukuba, 1-1-1 Tennodai, Tsukuba 305-8573, Japan*
*mhase@bk.tsukuba.ac.jp*

**Abstract:** We report on a direct generation of efficient and wavelength-tunable third-harmonic generation (THG) from ultrapure electronic-grade (EG) diamond crystals. Under an ultrafast infrared excitation at 1280 nm, the considerably high optical conversion efficiency of ~ 0.7% at a THG wavelength of 427 nm is obtained, and the THG signal can be tuned over ultra-broadband range from 420 to 730 nm. We argue that the THG efficiency is originating from minimum absorption loss and phase-matching conditions in EG diamond. Enhanced THG from EG diamond crystal represents a new paradigm for establishing efficient diamond-based frequency converters, quantum sensing, and quantum communications platforms.

KEYWORDS: Diamond, Third-harmonic generation, Femtosecond laser

## 1. Introduction

The generation of efficient and wavelength-tunable ultrashort coherent light sources in the visible spectral ranges is of enormous interest for the study of ultrafast processes, including time-resolved photodissociation dynamics of molecules [1]. Several established frequency conversion mechanisms can manage visible radiations, such as optical parametric amplification (OPA) [2], sum frequency generation, second harmonic generation (SHG) [3,4], and third harmonic generation (THG) [5,6], and so on. Among them, THG is a nonlinear optical (NLO) phenomenon that can reconcile three low-energy (longer wavelength) photons with frequency ω into a single high-energy (shorter wavelength) photon with three times of original frequency $\omega_{THG} = 3\omega$ [Fig. 1(a)]. In recent years, THG has gained extensive attention due to its unique optical properties and a broad range of applications in nonlinear scanning laser microscopy [7], UV-Visible spectroscopy [8], and telecommunications [9].

      On the other side, as illustrated in Fig. 1(b), a diamond crystal is composed of a single element of carbon, each carbon atom in the diamond crystal is linked to four other carbon atoms by strong covalent bonds: this rigid structure of a diamond makes it hardest substance among materials. Furthermore, diamond crystals have recently received a fair amount of attention and becoming prominent as promising optical material supported by its large bandgap energy (~5.5 eV) [10], and wide range of potential applications, including diamond-based laser [11–13], frequency convertors in terahertz region [14], photonic device technologies toward quantum photonic networks and quantum communications [15], quantum sensing [16], and single-photon sources [17].

      Over the past decades, substantial efforts have been devoted to the development of magnetic properties of diamond crystals [18,19], but little study has been done on the NLO properties of diamond crystals and THG-related experimental reports were limited to nanocrystalline diamonds [20,21]. Recently, we inquired into opto-magnetic effects in nitrogen-vacancy (NV) introduced diamond crystals to investigate the ultrafast dynamics of inverse Cotton–Mouton effect (ICME) induced by a spin ensemble from NV centers and discovered that the polarization dependence of the various NLO effects including ICME can be separated by considering the helicity dependence [22]. Furthermore, we and our colleagues explored the optical Kerr effect (OKE) in diamond crystals with NV color centers using an ultrashort pulse laser [23]. Kozak *et al* observed SHG and THG in the self-supporting nanocrystalline diamond membranes [20,21]. However, due to the atomic defects and imperfect phase-matched conditions at a surface, the obtained optical

conversion efficiency was limited to η = 5.6 × 10⁻⁸, and such weak TH signals hinder the practical applications of diamond crystals. Therefore, enhanced THG from diamond crystals offers novel "tricks" for light science, extending the functionality and applications of diamond optics and photonics.

In general, the key conditions in harmonic generation are nonlinear susceptibility and phase matching. The THG intensity becomes maximum when the phase-matching condition of $n_\omega = n_{3\omega}$ is satisfied, where $n_\omega$ and $n_{3\omega}$ are the refractive index for the fundamental and THG wavelengths, respectively [24,25]. In nonlinear optics, the third-order nonlinear polarization ($P$) of THG can be expressed by the following equation [25]:

$$\boldsymbol{P}_{THG}(3\omega) = \varepsilon_0 \chi^{(3)} \boldsymbol{E}(\omega)\boldsymbol{E}(\omega)\boldsymbol{E}(\omega) \qquad (1)$$

In this expression, $\varepsilon_0$ is the permittivity in vacuum, $\chi^{(3)}$ and $\boldsymbol{E}(\omega)$ are, respectively, the third-order nonlinear susceptibility and the incident electric field at frequency ω. The TH signal intensity $I(3\omega)$ can be expressed as [21,25]:

$$I(3\omega) = \frac{9\pi^2 L^2}{4\varepsilon_0^2 c^2 n_\omega^3 n_{3\omega} \lambda_\omega^2} |\chi^{(3)}(3\omega = \omega + \omega + \omega)|^2 \left|\frac{\sin(\Delta k L/2)}{\Delta k L/2}\right|^2 I(\omega)^3 \qquad (2)$$

where $L$ is the sample thickness, $c$ is speed of light, $\lambda_\omega$ is the wavelength of the pump beam, $\Delta k = k_{3\omega} - k_\omega$ is the phase mismatch, and $I(\omega)$ is the pump intensity. The TH signal intensity may saturate at a threshold pump intensity [6, 24].

In our most recent study, we have demonstrated SHG and cascaded THG (cTHG) from bulk diamond crystals using inversion symmetry-breaking by NV color center, in which the THG comprises two steps: initially, SH was generated by $\omega + \omega = \omega_{SHG}$: 2ω and then it is recombined with the fundamental wave to generate the third harmonic signals ($\omega_{SHG} + \omega = \omega_{THG}$: 3ω), and in this nonlinear process we argue both SHG and cTHG phase matching has to be satisfied [26,27]. We also investigated TH signals from centrosymmetric type-IIa pure bulk diamond crystals and found that the optical conversion efficiency of THG reached the η = 6.5±0.9×10⁻⁴ [26,27]. It is, therefore, an important task to generate enhanced harmonic signals from diamond crystals. Furthermore, electronic-grade (EG) single crystal diamond will become a nonlinear material with high third-order nonlinearity: significant enhancement of THG can be expected in the nonlinear optical response owing to ultra-pure crystal properties, whose impurity (nitrogen: [N] and boron: [B]) levels are [N] < 5 ppb and [B] < 1 ppb.

In this work, we present an efficient THG from an EG diamond crystal, thanks to the ultra-pure crystal structure and exquisitely phase matching. By irradiating IR femtosecond laser, centered at 1280 nm, which is far-bellow the band gap energy, we obtain strong visible TH signal at 427 nm, and the wavelength is tuned up to 730 nm by switching the pump wavelengths. The obtained optical conversion efficiency of THG from the ultra-pure EG diamond crystal is 2–3 and 6–7 orders of magnitude higher than conversion efficiency of type-IIa pure bulk (non EG) and nanocrystalline diamonds, respectively, and it's also comparable with 2D materials as well as other carbon-based materials like graphene [28,29].

## 2. Experimental details

The sample used was an <100> oriented single crystal EG diamond grown by the chemical vapor deposition (CVD) method from Element Six. The crystal size was 3.0 mm × 3.0 mm × 0.5 mm (thickness). Fig. 1(a) shows a schematic view of the operation designed to support a giant cubic nonlinear response for THG at visible wavelength regions, the deep red and blue regions indicate the incident pump beam at the IR wavelength and transmitted visible THG signals at 3ω, respectively.

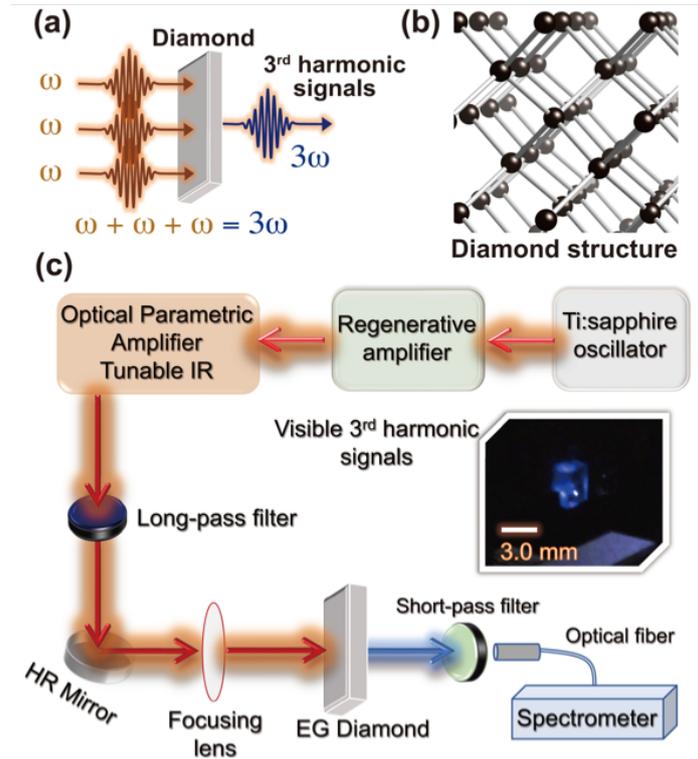

**Fig. 1**. (a) A schematic of third-harmonic generation (THG) from diamond crystals under the IR pump. (b) Atomic structure of the diamond crystal. (c) Optical layout of THG experiments: a mode-locked Ti:sapphire laser produces pulses at ~800 nm that is amplified and pump an optical parametric amplifier (OPA). For the THG experiments, we employed an OPA, wavelength-tunable between 1230 and 2190 nm. Visible blue emission from the surface of the EG diamond crystal is presented in the inset.

A simplified layout of the home-built optical spectroscopy setup used for the detection of nonlinear emissions from diamond crystals is illustrated in Fig. 1(c). The ultrafast laser pulses from a mode-locked Ti:sapphire laser (central wavelength at ~800 nm; repetition rate 80 MHz; pulse width 25 fs; and average power $P = 425$ mW) were amplified by a regenerative amplifier to obtain laser pulses at 100 kHz. The ultrafast laser pulse was used as a pump for the OPA, which generated 60 fs IR pulses with wavelength tunable between 1230 and 2190 nm.

The THG measurements are performed in air at room-temperature conditions with transmission optical geometry. The IR laser pulse was spectrally cleaned by a long-pass filter (FELH0950: OD5), and we employed a short pass filter (SHPF-25C-720: OD6) to separate the fundamental and third harmonic pulses; in addition to this, we used an additional visible bandpass filter (ET405/40x, 405 nm FWHM 40 nm: OD6 outside of the pass range) during the output power measurements. Thus, the use of highest contrast filters enabled us to obtain the THG signal. The excitation beam was focused by a focusing lens ($f = 50$ mm) to a beam radius of ~20 μm ($1/e^2$) at the EG diamond crystals. The emitted visible TH signals were separated with a filter and collected by a collimating objective lens to couple into an optical fiber and finally detected and integrated for 100 milliseconds by a UV-visible spectrometer. Note that we have observed a small deflection of the THG beam from the fundamental (pump) beam, which would be related to the phase-matching condition as discussed later. The pump and THG powers are measured before and after the sample, respectively, by using a microprocessor-based laser power/energy meter (Newport 1919-R), which can select the detecting wavelength for the calibration.

## 3. Experimental results and discussions

The emission spectra were recorded to figure out the nature of visible signals from the EG diamond sample. As illustrated in Fig. 2(a), under ultrafast IR illumination with center wavelength $\lambda_{(pump)} \approx 1280$ nm, the visible blue lights are simultaneously emitted from the EG diamond crystal. The nonlinear signals show relatively broadband spectra ranging from 410 to 450 nm and wavelength centered at $\lambda_{(THG)} \approx 427$ nm, which is ~1/3 of IR pump wavelength, being consistent with the relationship of the frequencies under THG. In addition, the emission spectra do not contain any other peaks, such as SHG, indicating that only the THG emission was obtained in the experiments. We note that the intense periodic spikes in emission spectra would come from interference effects at the front and back surfaces of the EG diamond crystal. Since breaking the spatial inversion symmetry occurs only at the surfaces of diamond crystals, cTHG from the SHG and the fundamental will be possible at the surfaces [26]. Thus, two cTHG beams from the front and back surfaces can interfere each other, and produce the periodic spikes [21,30].

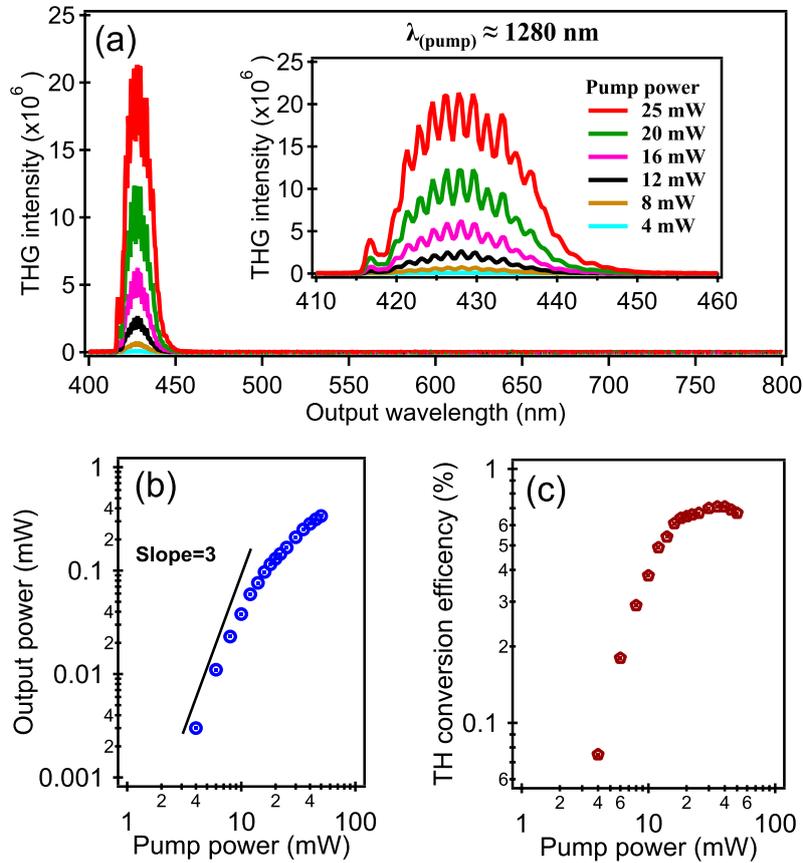

**Fig. 2.** (a) Dependence of blue third-harmonic generation (THG) spectra from the EG diamond crystal on the IR pump power. (b) The THG intensities as a function of pump power of IR laser. (c) THG conversion efficiency versus IR pump power for the EG diamond crystal.

To further examine if the blue light originated from a nonlinear optical process, THG, we recorded the blue light output power as a function of IR pump power at ≈1280 nm. The inserted enlarged view of Fig. 2(a) demonstrates the relation between blue light output power and IR pump power, which was varied from 4 to 50 mW (corresponding to the fluence from 3 to 38 mJ/cm$^2$). As presented in the double-logarithmic plot of the pump power versus blue light output power in Fig. 2(b), the output power clearly scales nonlinearly with the incident IR pump power and shows a slope close to three, which proves the frequency-tripling nature of the obtained blue light signals. Note that the relative error is less than 5%.

The THG optical conversion efficiency can be calculated by $\eta_{THG} = (P_{THG}/P_{pump}) \times 100$ from nonlinear media. As expected, the THG output power increases cubically as the IR pump power is increased, and as seen in Fig. 2(c) the THG conversion efficiency can reach a value of $\eta_{THG}$ ~0.7%, which is several orders of magnitude higher than any other THG efficiencies from nanocrystalline diamond and type-IIa pure bulk (non EG) diamond crystals. In addition, the obtained conversion efficiency is only one to two orders of magnitude smaller compared to the conventional 3rd-order nonlinear optical crystals [3,31]. The enhancement can be explained by better phase-matching in transmission measurements and negligible impurity absorption in ultrapure EG diamond crystal. We argue that there are several obvious routes to further improve this conversion efficiency, for example, by using CVD diamond without structural stresses (that can be characterized by the transmission pattern from crossed polarization measurements) or using nitrogen-free diamonds produced by high-pressure and high-temperature (HPHT) methods. Furthermore, as displayed in Fig. 2, the THG gives cubic intensity dependence, we obtained a maximum THG output power and maximum optical conversion efficiency of ~0.15 mW and ~0.7%, respectively, at the pump power of $P_{pump} = 20$ mW (corresponding peak excitation fluence of $F_{pump} \approx 15$ mJ/cm$^2$, the corresponding optical intensity is $I \approx 1.0$ TW/cm$^2$). However, above $P_{pump} > 20$ mW, saturation dominates and it reduces the detectable visible TH signals. We argue that the observed saturation would be due the strong-field regime beyond the perturbative nonlinear optics. In fact, the peak intensity of 2.55 TW/cm$^2$ ($P_{pump} = 50$ mW) in the present study matches with the condition in Ref.[32], in which impact ionization and optical field-induced tunneling were observed.

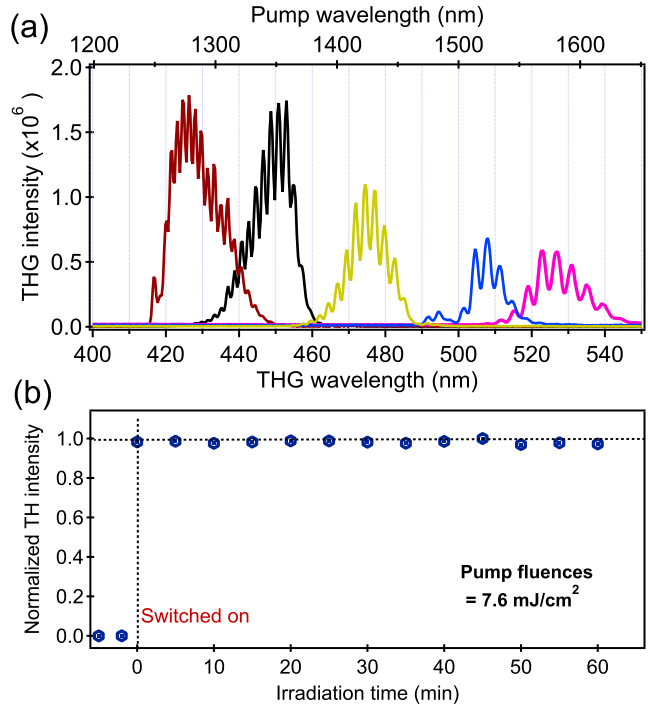

Fig. 3. (a) Nonlinear emission spectra of the EG diamond crystal. (b) Time dependence of TH emission intensity under ultrafast excitation wavelength at 1280 nm under constant optical fluences.

Figure 3(a) shows the THG spectra obtained for the pump wavelengths ranging from 1250 to 1590 nm (and also from 1930 to 2190 nm, data not shown). As displayed in Fig. 3(a), the THG response can be switched by the IR pump wavelength and the THG intensity becomes weaker when the excitation IR wavelength was detuned from the initial wavelength centered at 1280 nm. This observation is possibly due to the changes of the pulse duration and effective pump laser intensity (fluence). The first effect is because the single optical cycle ($T$) of the IR laser pulse generally becomes larger when the wavelength goes to longer, i.e., $T \approx 4.0$ fs for 1200 nm, while $T \approx 5.3$ fs for 1600 nm, thus it is plausible to get larger pulse width for the longer wavelength. For the second effect, we could get a larger laser spot size on the sample using the same focusing lens for the longer wavelength: this could decrease the effective pump intensity (fluence). Thus both two effects would decrease the THG intensity at the longer wavelength. Note that when the excitation IR wavelength is detuned from the initial wavelength centered at 1280 nm, the visibility of the spectral interference fringes decreases, possibly due to the modulation depth limitation by the spectrometer resolution ($\approx 2$ nm).

In addition to the excitation wavelength dependence, we investigated the time dependence of TH signals under the condition of a constant pump power 10 mW (optical fluences $F_{pump}$= 7.6 mJ/cm$^2$). As can be seen from Fig. 3(b), the time dependence of the TH emission intensity upon IR pump was nearly constant ($\pm 0.4\%$) after the light irradiation until 60 minutes, indicating a good photostability of the THG.

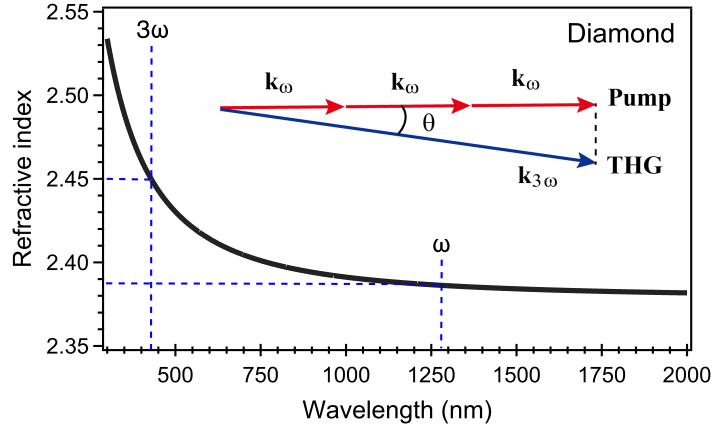

**Fig. 4**. Refractive index curve of ultrapure EG diamond crystals, calculated by the formula in Ref.[33]. The vertical dashed lines point the wavelength of the pump (1280 nm) and the corresponding THG (427 nm), respectively. The inset represents Cherenkov-type phase-matching conditions. $\boldsymbol{k}_\omega$ and $\boldsymbol{k}_{3\omega}$ represent the wavevector of the pump and THG, respectively, and $\theta$ is the angle between the two beams.

We now discuss a possible phase-matching mechanism in ultrapure EG diamond crystals. Figure 4 presents the refractive index curve of ultrapure diamond based on the dispersion function of Ref.[33]:

$$n^2(\lambda) = 1 + \frac{4.658\lambda^2}{\lambda^2 - 112.5^2}, \qquad (3)$$

where the wavelength $\lambda$ is expressed in *nm*. As shown in the refractive index curve in Fig. 4, the refractive index is *n*=2.45 at the THG wavelength, but *n*=2.39 at the fundamental wavelength, and the refractive index gives an almost constant value at the longer wavelength region. Since OKE has been observed for the EG diamond crystal from Element Six in Ref. [34], nonlinear refractive index ($n_2$) can change the refractive index according to $n = n_0 + n_2 I$, where $n_0$ is the bare refractive index and $I$ is the pump intensity [25].

We have estimated the value of $\Delta n = n_2 I$ to be $\approx 1.3 \times 10^{-3}$ using $n_2 \approx 0.5 \times 10^{-19}$ m$^2$/W from Almeida *et al.* [34] and $I = 2.55$ TW/cm$^2$ for our measurements, which is too small to obtain the phase-match condition, $n(3\omega) = n(\omega)$. Instead of this scenario, Cherenkov-type phase-matching can be possible when the refractive index change due to OKE occurs. In fact, we have observed a small deflection of the THG beam from the pump beam, whose angle was less than 10 degrees. In the Cherenkov-type phase matching, we have $n(3\omega)\cos\theta = n(\omega)$, as shown in the inset of Fig. 4. Then we obtain $\theta \approx 12.7$ degrees, which is nearly consistent with our experimental observations.

## 4. Conclusions

In conclusion, we have experimentally demonstrated efficient third harmonic generation from ultrapure electronic-grade diamond crystals at visible wavelengths with an optical conversion efficiency of η ~ 0.7%, which is approximately several orders of magnitude larger than the optical conversion efficiency of nanocrystalline and type-IIa pure bulk (non EG) diamonds [20,21,26]. We argue that the significant enhancement in the THG signal intensities could be originated from negligible photo absorption in the IR range and nearly phase-matching conditions. This THG conversion efficiency could be further improved, for example, by using CVD diamond crystals without structural stresses. Furthermore, the THG signals show good photostability and wide-band wavelength tuning characteristics. The efficient THG observed here will open a new practical and powerful avenue for developing diamond nonlinear optics and photonics, including diamond-based frequency converters, quantum sensing, and quantum communications platforms.

**Funding.** Core Research for Evolutional Science and Technology program of the Japan Science and Technology (Grant Number: JPMJCR1875).

**Present Address.**
⊥Department of Medicine, Washington University in St. Louis, 660 S. Euclid Ave., St. Louis, 63110-1010, USA.

**Disclosures.** The authors declare no conflicts of interest.

**Data availability statement.** Data underlying the results presented in this paper are not publicly available at this time but may be obtained from the authors upon reasonable request.